\documentclass[preprint]{aastex631}

\accepted{August 23, 2021}

\shorttitle{Altered State of eta Car}
\shortauthors{Martin, et al.}

\graphicspath{{./}{figures/}}
\begin{document}

\title{The Altered State of $\eta$ Carinae: HST's Photometric Record 1998--2021}
\altaffiliation{Based on observations made with the NASA/ESA Hubble Space Telescope, obtained from the data archive at the Space Telescope Science Institute. Support for this work was provided by NASA through grant number GO-16176 from the Space Telescope Science Institute.  STScI is operated by the Association of Universities for Research in Astronomy, Inc. under NASA contract NAS 5-26555.}

\correspondingauthor{John C. Martin}
\email{jmart5@uis.edu}

\author[0000-0002-0245-508X]{John C. Martin}
\affiliation{University of Illinois Springfield}

\author[0000-0003-0221-788X]{Kris Davidson}
\affiliation{University of Minnesota}

\author[0000-0003-1720-9807]{Roberta M. Humphreys}
\affiliation{University of Minnesota}

\author{Kazunori Ishibashi}
\affiliation{University of Nagoya}

\begin{abstract}
Hubble Space Telescope photometry of $\eta$ Carinae spans 23 years,
including five spectroscopic events.  The rapid brightening rate
decreased after 2010, and the spectroscopic events in 2014 and 2020 had light curves
different from their predecessors.  Together with other indicators,
these developments probably  foretell the conclusion of $\eta$ Car's
change of state.
\end{abstract}

%% https://urldefense.com/v3/__https://astrothesaurus.org__;!!DZ3fjg!rJL29Je14OEioVPJw0witWkXbGJO71T-QDdqhynwOxZ4usVduq6JoGZBB_zn96E$ 

\section{}
Since the 1990's, the central star of $\eta$ Car has brightened 
by several magnitudes at UV-to-red wavelengths  
\citep{2006AJ....132.2717M,2010AJ....139.2056M,1999AJ....118.1777D}  
without much change in luminosity \citep{2019A&A...630L...6M}.  
The most likely cause is a rapid decrease of circumstellar extinction,  
consistent with spectral changes that indicate a 
decline in the mass outflow rate
\citep{2010ApJ...717L..22M,2012ApJ...751...73M,2018ApJ...858..109D}.   
But the rate of brightening has diminished since 2010,  and 
spectroscopic evidence of its companion star near periastron has also changed. 
These facts suggest that $\eta$ Car has nearly recovered from 
its 1830-1860 Great Eruption.  The Hubble Space Telescope (HST) has 
been essential for this problem (see below) but is not expected to 
produce significantly more data for this topic.  Hence it is appropriate 
now to summarize the HST photometric record from 1998 to 2021.  

For general information about $\eta$ Car, see reviews by many 
authors in \citet{2012ASSL..384.....D}, hereinafter ``Revs2012.'' 
This is the only supernova impostor that is close enough and 
recent enough to provide abundant clues to the post-eruptive structure.   
HST's spatial resolution is essential, because ground-based data are 
severely contaminated by bright ejecta within $0.5\arcsec$ of the star  
(\citealt{2006AJ....132.2717M,2010AJ....139.2056M}; 
Fig.\ 1 in \citealt{2015ApJ...801L..15D}).  
Over the past 23 years HST/WFPC2, HST/ACS, and HST/STIS have 
recorded the star's apparent brightening.  The synthetic STIS 
photometry, being immune to variable atmospheric seeing, is the 
most homogeneous time series ever done for $\eta$ Car.  

Figure 1 shows the main results expressed as STMAG values (see caption), 
with no attempt to correct for any extinction.   
F250W and F330W fluxes are equivalent to medium-bandwidth UV 
photometry at $\lambda \approx$ 250 nm and 330 nm, see    
\citet{2006AJ....132.2717M,2010AJ....139.2056M}.  
They include absorption and emission lines.  The blue 477 nm and 
red 690 nm fluxes, however, represent narrow continuum intervals 
with no substantial absorption or emission features 
(cf.\ figures in \citealt{2015ApJ...801L..15D}).  They used a  
$0.1\arcsec \, \times \, 0.2\arcsec$   
sampling aperture, with crude throughput corrections of $1.5\times$ 
at 477 nm and $1.7\times$  at 690 nm.  These correction factors are 
uncertain by perhaps $\pm$10\% because the central object is not 
a point source;  but random  errors in the brightness {\em variations} 
were usually less than 4\%.   The rapid up-and-down fluctuations in 2014 
and 2020 appear to be real, not caused  by imperfect pointing, since the 
recorded spacecraft jitter and other details were similar to the pre-2004 
observations.

Two different phenomena are obvious in Fig.\ 1:  progressive brightening, 
and brief minima at 5.54-year intervals.  The brightening trend   
probably indicates a declining circumstellar extinction, due to a 
lessening rate of dust formation.    
According to standard temperature arguments, dust grains should form at 
$r \, \sim$ 100--300 AU in the star's mass outflow, in material that 
was ejected 1--5 years earlier.  This explanation is consistent with 
spectroscopic evidence for a decreasing mass loss rate 
(\citealt{2018ApJ...858..109D} and many refs.\ therein), and no satisfactory  
alternative has been  proposed.  The apparent color measured from 
the star's visual continuum has changed 
remarkably little, but this may be acceptable since $\eta$ Car notoriously 
forms very large dust grains (Revs2012).  In any case, if we fit a smooth upper 
envelope to each light curve in Fig.\ 1, the brightening rate averaged 
0.18 magn/y in 2000--2010 but only 0.08 magn/y in 2010--2020.  

Abrupt temporary changes in 1998, 2003, 2009, 2014, and 2020 coincide   
with ``spectroscopic events'' when high excitation lines and other
features in the spectrum react to the companion star's periastron  
passages (Revs2012).   The primary star was {\em not} eclipsed at those times.  
Deep minima in the 250 nm waveband are caused by a forest of \ion{Fe}{2} 
UV absorption lines, and the 330 nm brightness is largely Balmer 
continuum, but otherwise the short-term brightness changes have not yet 
been explained.  Size and time scales (several AU, several weeks) 
were far too small for  dust formation.  
Conceivably the line-of-sight extinction might vary rapidly if dust
destruction by the  star's radiation competes with grain formation, a
subtle idea which we hope to discuss in a later paper.  Alternatively, if  
the photosphere is located 
in an opaque outflow, it can be perturbed by tidal effects as well as 
varying photoionization by the O4-type secondary star. 

Even without a clear theory, however, Fig.\ 1 shows that the 2014 and 2020 
events differed from their predecessors.  At wavelengths longer than 
300 nm, the light curve for the 2003 and 2009 events resembled a well-defined  
square root symbol ($\surd$) preceded by relatively smooth 
two-year declines of roughly 0.3 magnitude and with a smooth 
rapid increasing flux after minimum.    In 2014 and especially 2020, 
however, brightness spikes occurred several weeks before minimum and 
and then the post-minimum brightening had a much smaller amplitude than 
before. (Details will appear in a later paper.)  These differences 
arose soon after the observed spectroscopic change 
of state before 2009 or 2010  
\citep{2010ApJ...717L..22M,2015ApJ...801L..15D,2015A&A...578A.122M}.     

Indeed, our HST/STIS data show that the 2020 event spectroscopically 
resembled  its predecessor in 2014, but greatly differed from earlier  
instances (analysis currently in progress).  Temporary shortages of 
ionizing UV, from the secondary, defined the periastron events 
before 2010, but not since. 
In the simplest hypothesis proposed so far, accretion from the primary 
wind quashed the secondary star's EUV near periastron before 2010  
(\citealt{2009NewA...14...11K} and earlier refs.\ cited therein), 
but later the declining primary wind density could not provide  
sufficient accretion \citep{2015ApJ...801L..15D}.   

To some extent we can foresee $\eta$ Car's near-future appearance.  
Its continuum brightness has lately risen to $V \approx 4.8$, 
not including emission lines and the Homunculus ejecta-nebula.
If the rate of brightening continues to follow the upper-envelope 
curve implied in Fig.\ 1, it will level off near $V \approx 4.0$ 
about two decades from now -- close to the  
brightness that Halley recorded in 1677 
\citep{1903AnCap...9...75I,2004JAD....10....6F}.  
Meanwhile, with 
declining gas and dust densities, the hot secondary star will begin 
to photoionize the Homunculus.  It will then resemble a titanic planetary 
nebula with a surface brightness far more intense 
than any familiar planetary nebula.  
\explain{Rightly so MS Exchange identifies email fro arXiv moderators as "junk" but it is junk we have to listen to.}
\explain{arXiv moderators are probably insiting we include a table because its the hardest most time consuming part of writing LaTex.}

\begin{figure}
\figurenum{1}
\label{fig1}
\plotone{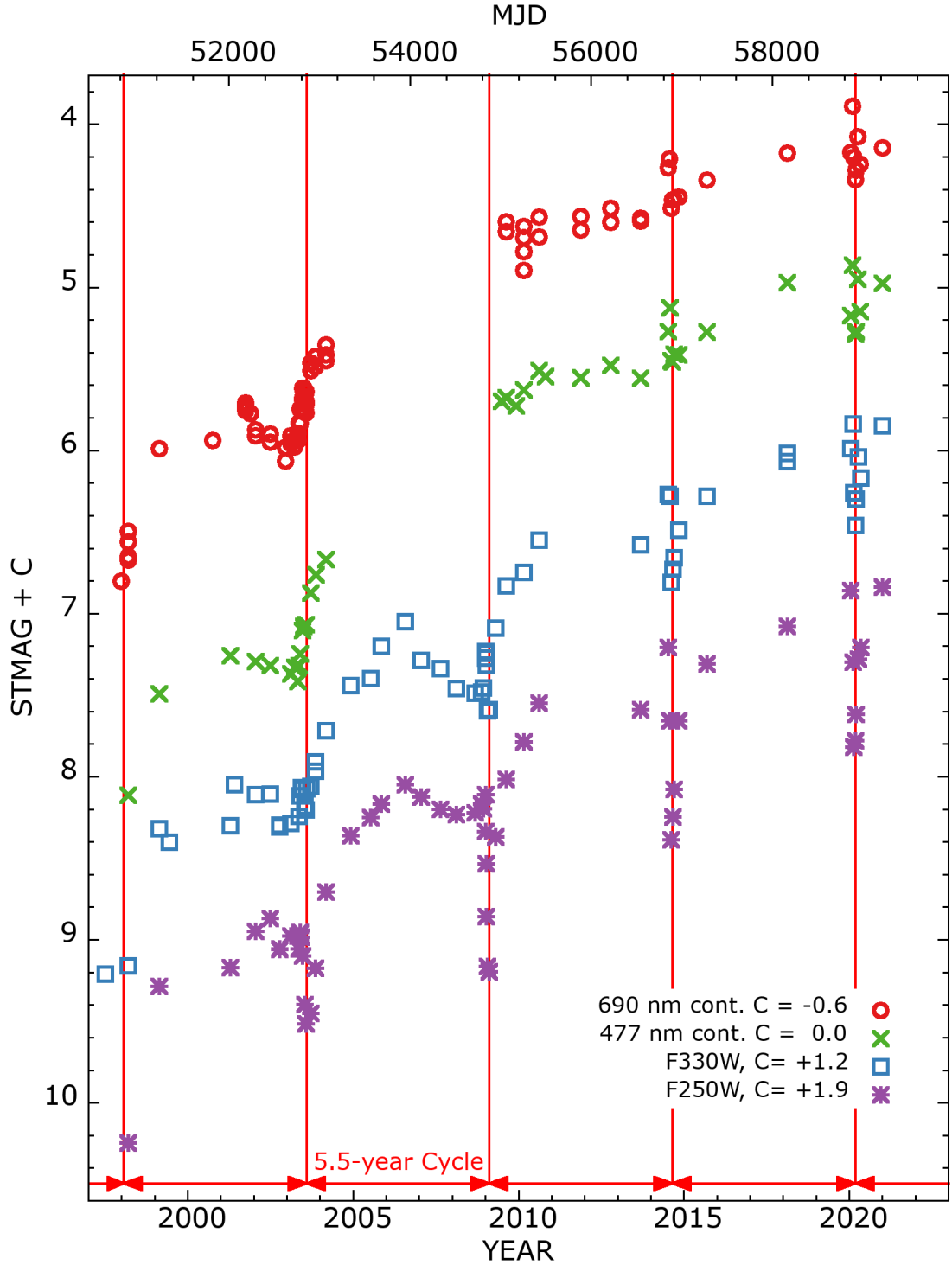}

\caption{Apparent brightness of the central star measured by HST, expressed as
   STMAG $= \; -2.5 \, {\log}_{10} \, (f_{\lambda}/f_0)$ where 
   $f_0 = 3.63 \times 10^{-9}$ erg cm$^{-2}$ s$^{-1}$ {\AA}$^{-1}$. 
   Most points are synthesized from STIS calibrated spectra, while ACS 
   and WFPC2 data fill the 2005--2009 hiatus in STIS coverage.    
   For methods and details, see \citet{2010AJ....139.2056M}.}

\end{figure}

\startlongtable
\begin{deluxetable}{ccccc}
\tablecolumns{5}
\tablecaption{Table of the magnitudes plotted in Figure \ref{fig1}.\label{tab1}}
\tablehead{\colhead{Epoch} & \colhead{690 nm (mag)}&\colhead{477 nm (mag)}&\colhead{F330W (mag)}&\colhead{ F250W (mag)}}
\startdata
1994.28&&8.22\\      
1997.43&&7.77\\      
1997.52&&8.01\\      
1998.00&7.44\\                
1998.21&7.20&7.99&7.96&8.35\\ 
1998.21&7.13\\                
1998.21&7.31\\                
1998.21&7.29 \\               
1999.14&6.63&7.377&7.12&7.39\\ 
1999.44&&7.20\\      
2000.77&6.56\\                
2001.29&&7.13&7.10&7.27\\ 
2001.42&&6.85\\
2001.75&6.38\\                
2001.75&6.38\\                
2001.75&6.33\\                
2001.75&6.35\\                
2001.90&6.40\\                
2002.05&6.50&7.17&6.91&7.05\\ 
2002.05&6.53\\                
2002.51&6.52\\                
2002.51&6.58&7.19&6.91&6.97\\ 
2002.78&&&7.11&7.16\\ 
2002.78&&&7.16\\ 
2002.78&&&7.10\\      
2002.96&6.60\\                
2002.96&6.68\\                
2003.12&6.54&&7.09&7.08\\ 
2003.12&6.58\\                
2003.12&6.57&7.24&7.09&7.08 \\
2003.24&6.60&7.20\\           
2003.34&6.52\\                
2003.34&6.56&7.29\\           
2003.37&&7.04&7.16\\ 
2003.38&6.46&7.19\\           
2003.41&6.38&7.13&6.92&7.06\\ 
2003.42&6.46\\                
2003.42&6.37\\                
2003.45&&&6.87&7.09\\ 
2003.47&6.31&&6.88&7.20\\ 
2003.47&6.25\\                
2003.47&6.30\\                
2003.48&&6.98\\           
2003.51&6.39&6.95\\           
2003.51&6.24\\                
2003.51&6.29\\                
2003.50&&&6.98&7.50\\ 
2003.58&6.32&6.94&7.01&7.62\\ 
2003.58&6.39\\                
2003.58&6.33\\                
2003.58&6.26\\                
2003.60&&6.88\\      
2003.72&6.09&6.75&6.86&7.55\\ 
2003.72&6.13\\                
2003.87&&6.71&7.28\\ 
2003.87&&6.77\\      
2003.88&6.05&6.64\\           
2003.88&6.11\\                
2004.18&6.07&6.55&6.52&6.81\\ 
2004.18&5.99\\                
2004.18&6.05\\                
2004.93&&&6.24&6.46\\ 
2005.53&&&6.20&6.35\\
2005.85&&&6.00&6.27\\ 
2006.59&&&5.85&6.15\\ 
2007.06&&&6.09&6.23\\ 
2007.64&&&6.14&6.30\\ 
2008.12&&&6.26&6.33\\ 
2008.69&&&6.29&6.32\\ 
2008.88&&&6.28&6.27\\ 
2008.94&&&6.26&6.30\\ 
2009.01&&&6.07&6.21\\ 
2009.02&&&6.04&6.44\\ 
2009.03&&&6.04&6.63\\ 
2009.04&&&6.12&6.96\\ 
2009.07&&&6.40&7.26\\
2009.11&&&6.39&7.30\\ 
2009.33&&&5.89&6.47\\ 
2009.49&5.58\\           
2009.63&5.26&5.55&5.63&6.12\\ 
2009.63&5.20\\                
2009.93&&5.60\\           
2010.17&5.30&5.50&5.55&5.89\\ 
2010.17&5.23\\                
2010.17&5.49\\                
2010.17&5.38\\                
2010.63&5.29&5.39&5.35&5.65\\ 
2010.63&5.17\\                
2010.82&&5.42\\           
2011.89&5.25&5.43\\           
2011.89&5.17\\               
2012.80&5.22&5.35\\           
2012.80&5.14\\                
2013.70&5.21&5.43&5.38&5.69\\ 
2013.70&5.19\\                
2014.53&4.87&5.15&5.07&5.31\\ 
2014.58&4.82&5.01&5.08&5.76\\ 
2014.62&5.11&5.33&5.61&6.49\\
2014.66&5.06&5.34&5.53&6.35\\ 
2014.71&5.06&5.29&5.46&6.18\\ 
2014.86&5.05&5.29&5.29&5.76\\
2015.70&4.94&5.15&5.08&5.41\\ 
2018.14&4.77&4.84&4.87&5.18\\ 
2018.14&&4.82\\      
2020.05&4.77&5.04&4.79&4.96\\ 
2020.11&4.49&4.74\\           
2020.12&&4.64&5.40\\ 
2020.15&4.80&&5.06&5.92\\ 
2021.01&4.73&4.84&4.65&4.94\\ 
2020.20&4.92&5.16&5.26&5.88\\ 
2020.22&4.87&5.14&5.10&5.72\\ 
2020.27&4.67&4.82&4.84&5.38\\ 
2020.35&4.83&5.02&4.97&5.31\\ 
\enddata
\end{deluxetable}


\begin{thebibliography}{}

%%%  Some of these are commented out because they no longer appear in the text.

%%%  \bibitem[Damineli(1996)]{1996ApJ...460L..49D} Damineli, A.\ 1996, \apjl, 460, L49. doi:10.1086/309961
	
\bibitem[Davidson et al.(1995)]{1995AJ....109.1784D} Davidson, K., Ebbets, D., Weigelt, G., et al.\ 1995, \aj, 109, 1784. doi:10.1086/117408

\bibitem[Davidson et al.(1999)]{1999AJ....118.1777D} Davidson, K., Gull, T.~R., Humphreys, R.~M., et al.\ 1999, \aj, 118, 1777. doi:10.1086/301063

\bibitem[Davidson \& Humphreys(2012)]{2012ASSL..384.....D} Davidson, K. \& Humphreys, R.~M., eds., 2012, Eta Carinae and the Supernova Impostors: , Astrophysics and Space Science Library, Volume 384. ISBN 978-1-4614-2274-7. Springer Science+Business Media, LLC, 2012. doi:10.1007/978-1-4614-2275-4

%%%  \bibitem[Davidson et al.(2017)]{2017RNAAS...1....6D} Davidson, K., Ishibashi, K., \& Martin, J.~C.\ 2017, Research Notes of the American Astronomical Society, 1, 6. doi:10.3847/2515-5172/aa96b3

\bibitem[Davidson et al.(2018)]{2018ApJ...858..109D} Davidson, K., Ishibashi, K., Martin, J.~C., et al.\ 2018, \apj, 858, 109. doi:10.3847/1538-4357/aabdef

\bibitem[Davidson et al.(2015)]{2015ApJ...801L..15D} Davidson, K., Mehner, A., Humphreys, R.~M., et al.\ 2015, \apjl, 801, L15. doi:10.1088/2041-8205/801/1/L15

\bibitem[Frew(2004)]{2004JAD....10....6F} Frew, D.~J.\ 2004, Journal of Astronomical Data, 10, 6 

%%%%  \bibitem[Feast et al.(2001)]{2001MNRAS.322..741F} Feast, M., Whitelock, P., \& Marang, F.\ 2001, \mnras, 322, 741. doi:10.1046/j.1365-8711.2001.04163.x

%%%%  \bibitem[Humphreys \& Koppelman(2005)]{2005ASPC..332..159H} Humphreys, R.~M. \& Koppelman, M.\ 2005, The Fate of the Most Massive Stars, 332, 161

\bibitem[Innes \& Kapteyn(1903)]{1903AnCap...9...75I} Innes, R.~T.~A. \& Kapteyn, J.~C.\ 1903, Annals of the Cape Observatory, 9, 2.75

\bibitem[Kashi \& Soker(2009)]{2009NewA...14...11K} Kashi, A. \& Soker, N.\ 2009,New Astronomy, 14, 11   doi:10.1016/j.newast.2008.04.003  

\bibitem[Martin et al.(2006)]{2006AJ....132.2717M} Martin, J.~C., Davidson, K., \& Koppelman, M.~D.\ 2006, \aj, 132, 2717. doi:10.1086/508933

\bibitem[Martin et al.(2010)]{2010AJ....139.2056M} Martin, J.~C., Davidson, K., Humphreys, R.~M., et al.\ 2010, \aj, 139, 2056. doi:10.1088/0004-6256/139/5/2056

\bibitem[Mehner et al.(2019)]{2019A&A...630L...6M} Mehner, A., de Wit, W.-J., Asmus, D., et al.\ 2019, \aap, 630, L6. doi:10.1051/0004-6361/201936277

\bibitem[Mehner et al.(2010)]{2010ApJ...717L..22M} Mehner, A., Davidson, K., Humphreys, R.~M., et al.\ 2010, \apjl, 717, L22. doi:10.1088/2041-8205/717/1/L22

%%%%  \bibitem[Mehner et al.(2010)]{2010ApJ...710..729M} Mehner, A., Davidson, K., Ferland, G.~J., et al.\ 2010, \apj, 710, 729. doi:10.1088/0004-637X/710/1/729

\bibitem[Mehner et al.(2012)]{2012ApJ...751...73M} Mehner, A., Davidson, K., Humphreys, R.~M., et al.\ 2012, \apj, 751, 73. doi:10.1088/0004-637X/751/1/73

\bibitem[Mehner et al.(2015)]{2015A&A...578A.122M} Mehner, A., Davidson,K. Humphreys, R.~M., et al.\ 2015, \aap, 578, A22.  doi:10.1051/0004-6361/201425522  

%%%%  \bibitem[Parkin et al.(2009)]{2009MNRAS.394.1758P} Parkin, E.~R., Pittard, J.~M., Corcoran, M.~F., et al.\ 2009, \mnras, 394, 1758. doi:10.1111/j.1365-2966.2009.14475.x

%%%%  \bibitem[Remmen et al.(2013)]{2013ApJ...773...27R} Remmen, G.~N., Davidson, K., \& Mehner, A.\ 2013, \apj, 773, 27. doi:10.1088/0004-637X/773/1/27

%%%%  \bibitem[Soker(2005)]{2005ApJ...635..540S} Soker, N.\ 2005, \apj, 635, 540. doi:10.1086/497389

%%%%  \bibitem[Zethson et al.(2012)]{2012A\&A...540A.133Z} Zethson, T., Johansson, S., Hartman, H., et al.\ 2012, \aap, 540, A133. doi:10.1051/0004-6361/201116696
\end{thebibliography}
\end{document}